\documentclass[prd,aps,amsmath,amssymb,showpacs]{revtex4}
\usepackage{graphicx}
\usepackage{epsfig,rotate,latexsym}
\begin{document}
\title{Analyzing flow anisotropies with excursion sets in 
relativistic heavy-ion collisions}
\author{Ranjita K. Mohapatra}
\email {ranjita@iopb.res.in}
\author{P. S. Saumia}
\email {saumia@iopb.res.in}
\author{Ajit M. Srivastava}
\email{ajit@iopb.res.in}
\affiliation{Institute of Physics, Sachivalaya Marg, 
Bhubaneswar 751005, India}
%
%

\begin{abstract}
 
We show that flow anisotropies in relativistic heavy-ion collisions 
can be analyzed using a certain technique of shape analysis of 
excursion sets recently proposed by us for CMBR fluctuations to 
investigate anisotropic expansion history of the universe. The 
technique analyzes shapes (sizes) of patches above (below) certain
threshold value for transverse energy/particle number (the
excursion sets) as a  function of the azimuthal angle and rapidity. 
Modeling flow by imparting extra anisotropic momentum to the momentum 
distribution of particles from HIJING, we compare the resulting 
distributions for excursion sets at two different azimuthal angles. 
Angles with maximum difference in the two distributions identify 
the event plane, and the magnitude of difference in the two distributions 
relates to the magnitude of momentum anisotropy, i.e. elliptic flow. 
\end{abstract}

\maketitle

 Analysis of anisotropies of flow in relativistic heavy-ion collision
experiments (RHICE) has provided valuable information regarding the
properties of matter produced during early transient stages in
these collisions \cite{flow0}. Early investigations focused on the
elliptic flow relating to the 2nd Fourier coefficient of momentum
anisotropy as a function of azimuthal angle, as well as couple of other 
flow coefficients. Drawing correspondence of the underlying physics of 
fluctuations between RHICE and the Cosmic Microwave Background Radiation 
(CMBR) anisotropies, 
it was pointed out by us in ref.\cite{cmbhic} that a wealth of information
is contained in plots of root-mean-square values of flow coefficients
$v_n^{rms}$ for a large range of values of $n$, from 1 to 30-40. It was 
also proposed in \cite{cmbhic} that  the presence of superhorizon
fluctuations can lead to the suppression of superhorizon modes, as well as
the existence of acoustic oscillations in plots of $v_n^{rms}$ in RHICE, 
similar to the celebrated acoustic peaks of CMBR arising from inflationary  
density fluctuations. This suppression of superhorizon modes in RHICE 
may have been seen in the analysis reported by Sorenson for RHIC data 
\cite{srnsn}.

 This very useful interplay of physics from CMBR to RHICE was further
explored in ref.\cite{mhd} following the effects of magnetic field
on the power spectrum of CMBR anisotropies \cite{cmbrB}. Utilizing 
that insight, it was proposed in \cite{mhd} that presence of strong 
magnetic field in the initial stages of a non-central collision in RHICE 
can affect flow anisotropies, for example, it can lead to enhancement of 
flow coefficients. In particular, it could lead to enhancement in the 
elliptic flow coefficient $v_2$ by almost 30 \% (which may allow
for a higher value of viscosity). A different analysis in 
ref.\cite{tuchin} in the strong field limit further explored 
these effects and gave similar results for the enhancement of $v_2$.

 In this paper we explore another common feature between the analysis of
CMBR and the flow analysis in RHICE. This is based on a technique recently 
proposed by us for investigating any anisotropic expansion stage in the 
history of the universe by analyzing the shapes of patches of CMBR 
fluctuations (the excursion sets) at the surface of last scattering 
\cite{expn}. The main idea underlying this analysis \cite{expn} was the 
following. If the density perturbations generated initially are 
statistically isotropic, then they will get deformed in a specific
direction if the universe ever went through an anisotropic expansion.
This (statistically averaged)  deformation can be detected by a shape
analysis of fluctuation patches of CMBR. An anisotropic expansion of the 
universe has very strong similarities with the anisotropic flow expected 
in RHICE. The situation of anisotropic expansion of the universe 
is directly realized in the form of elliptic flow for non-central 
events.  Flow anisotropies are generally present in every event in RHICE 
and all values of $v_n^{rms}$ are non-zero in general due to the presence 
of fluctuations of different scales even in the central collisions 
\cite{cmbhic}. This suggests that the shape analysis of excursion sets 
used in ref.\cite{expn} should provide useful information for flow 
anisotropies in RHICE as well, especially for the elliptic flow. We
will see below that this is indeed true. We consider fluctuations
of transverse energy or particle number as a  function of the 
azimuthal angle and rapidity. This is like CMBR fluctuations
at the last scattering surface. Then by considering fluctuations
above/below a certain threshold value, we generate the excursion
sets of fluctuations, we call them simply as fluctuation patches.
To see the possibility of detecting anisotropic flow, we model 
flow by imparting extra anisotropic momentum to the momentum distribution 
of particles obtained from HIJING \cite{hijing}. We then analyze the 
shape/size distribution of the fluctuation patches at two different 
azimuthal angle. We will see that azimuthal angles with maximum 
difference in the two distributions identify the event plane, and the 
magnitude of difference relates to the magnitude of momentum anisotropy, 
i.e. elliptic flow. 

 For analyzing anisotropic expansion history of the universe in \cite{expn},
we had examined the size distribution of patches on the surface of last 
scattering (which is a two sphere $S^2$) along $\theta$ and $\phi$ directions.
This will be different for RHICE. Since our interest is in probing
flow anisotropy, say the elliptic flow in a non-central event, we would like
to compare the shape/size distribution of fluctuations along the X axis
with that along the Y axis, where the X and the Y axis correspond to
the event plane. As we mentioned, we model anisotropic flow in HIJING
by multiplying a factor $f_{p_x}$ to the momentum $P_x$ of every
final state particle. For central events $f_{p_x}$ is taken to be 1
while for non-central events $f_{p_x}$ is taken to vary from 1.1 to 1.4
representing the momentum anisotropy resulting from elliptic flow.
Precise value of momentum anisotropy represented by $f_{p_x}$ is not
an issue here (though the values used here are of right order of
magnitude, say, $f_{p_x}$ = 1.2 implying about 20 \% anisotropy).
Our intention is only to show the systematic pattern in which the
difference between the size distributions  of excursion sets at 
different azimuthal angle probe the event plane and strength of
elliptic flow. A quantitative correspondence can only be established
by incorporating our analysis in  hydrodynamical simulations and
we hope to report such an analysis in a future work.
 
  With the (anisotropic) momentum distribution of particles from HIJING,
we collect particle numbers as well as transverse energy in different
$\phi$ and $y$ bins where $\phi$ is the azimuthal angle in the event plane
and $y$ is the rapidity. We use 100 bins for full $2\pi$ range of $\phi$ 
and 100 bins for the full range of $y$. We have also considered smaller range
of rapidity $y$ and the results are similar. Full range of $y$ gives
better statistics. To exclude very large values of $y$ we use a lower
cutoff for $P_T \simeq $ 150 MeV. For each event thus we get the
particle number/transverse energy as a function of $\phi$ and
$y$. By subtracting the average values of these quantities we get 
particle number fluctuations/transverse energy fluctuations as a 
function of $\phi$ and $\eta$. We then consider a threshold value
for these fluctuations and consider only those $\phi,y$ bins where
particle number or transverse energy fluctuations is above 
(or below) this threshold value. This provides us with the excursion 
sets for the relevant fluctuations. Fig.1 shows the excursion sets
for transverse energy fluctuations obtained from HIJING for a single
Au-Au collision event at $\sqrt{s_{NN}}$ = 2 TeV. (Here, we
will be presenting results for transverse energy fluctuations. 
results for number fluctuations are very similar to these, hence
we do not discuss them separately.) We consider here non-central events 
with the impact parameter lying between $b = 3 - 4$ fm. Here we
have used $f_{p_x}$ = 1.2 (for modification of the momenta of
particles from HIJING) and fluctuations above 0.2 $\times$
maximum fluctuation have been included to generate the excursion sets.
We use here large collision energy as relevant for LHC. The size distributions
for small number of particles have large statistical errors. For the
case of RHIC energies one will need to combine much larger number
of events to control statistical errors.

\begin{figure}
\begin{center}
\leavevmode
\epsfysize=5truecm \vbox{\epsfbox{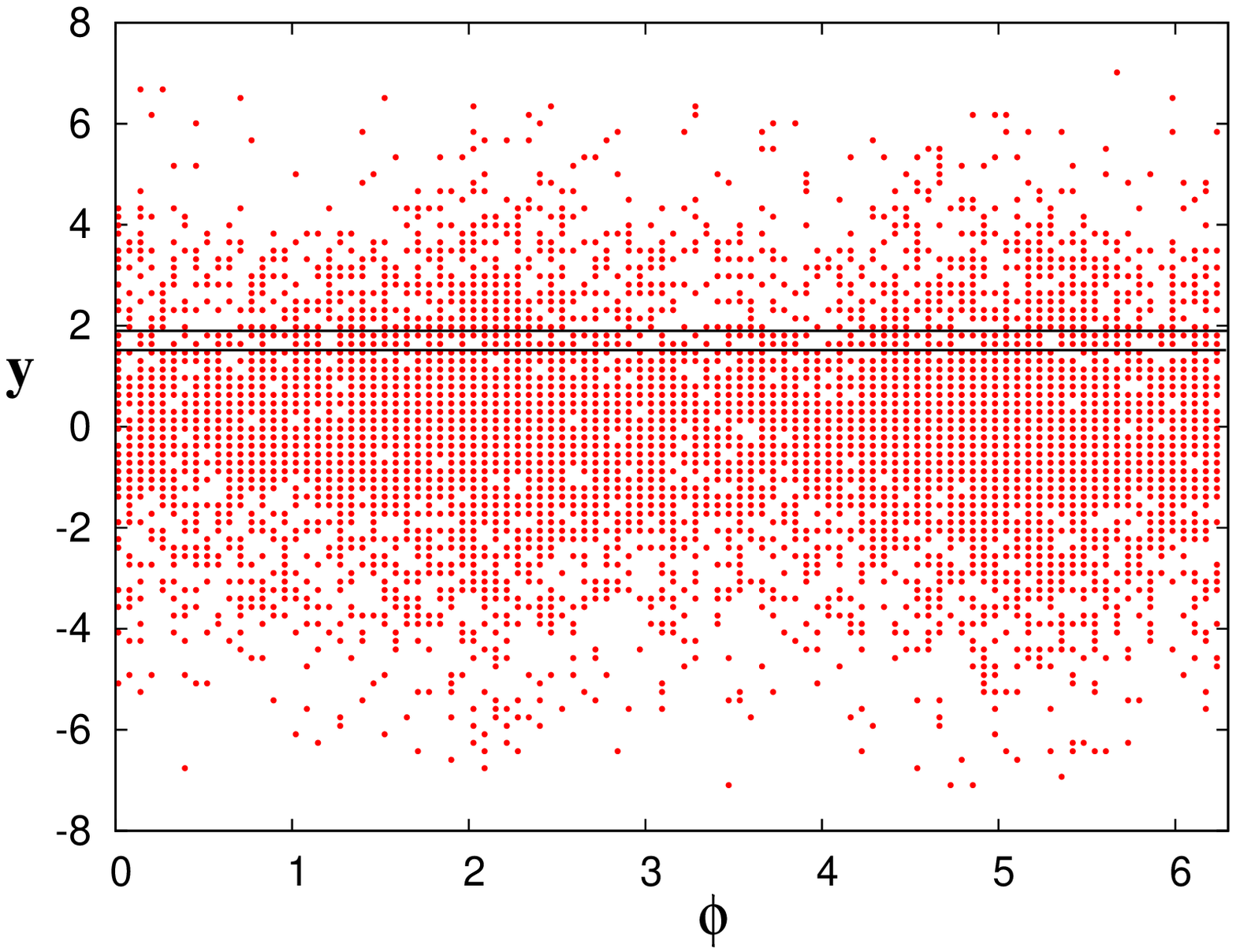}}
\end{center}
\caption{}{Plot of excursion sets (fluctuation patches) in the
azimuthal angle $\phi$ (in radians) and the rapidity $y$ space.
Horizontal solid lines show the slicing used for widths of filled patches.}
\label{Fig1}
\end{figure}

What we want to know is the average width of the fluctuations along 
$\phi$  direction near $\phi$ = 0 and compare it with the similar 
average width near $\phi = \pi/2$. This will probe the elliptic flow.
(Clearly, the same result will be obtained by comparing distributions 
at $\phi=\pi$ and $\phi=3\pi/2$.) We determine the distribution of
widths along the $\phi$ direction in a 45$^\circ$ wide arc centered
at $\phi = 0$ and compare it with the distribution of widths (again,
in a 45$^\circ$ wide arc) centered at $\phi = \pi/2$. To calculate 
these distributions we proceed as follows. We divide the entire 
region shown in Fig.1 (with $y$ varying within full range, and $\phi$ 
ranging from 0 to $360^\circ$) into thin slices (with a rapidity width 
of 0.18) along the $\phi$ direction, as shown by horizontal lines in
Fig.1. Using these slices, we determine the  $\phi$ widths of various 
filled  patches. We then plot the frequency distributions (histograms) of 
the widths of the intersections of all the patches with these slices at 
$\phi = 0$ and at $\phi = \pi/2$ (within 45$^\circ$ arcs). For the isotropic 
expansion case, without any flow anisotropy, we expect these two
histograms to almost overlap. For anisotropic expansion case we expect
that fluctuation patches will be stretched more at $\phi = \pi/2$ than at
$\phi = 0$, as system expands faster along $\phi = 0$ (and $\phi = \pi$)
directions for a non-central collision. Any relative difference, between 
the two histograms will, therefore, imply the presence of an anisotropy
of expansion (apart from the possibility of any initial anisotropy,
we will discuss it below).

Fig. 2a shows the frequency distributions (histograms) of the widths of the 
intersections of the slices with the filled patches (excursion sets) at 
$\phi = 0$ (solid curve) and $\phi = \pi/2$ (dashed curve) for the case 
with $f_{p_x}$ = 1 (i.e. no momentum anisotropy. The horizontal axis 
corresponds to the widths of the slices (in radians) along $\phi$,  with 
histogram bin having width of 0.05 radians. The vertical axis gives the 
frequency $N$ of the occurrence of the respective widths in all the 
slicings  of excursion sets (such as in Fig.1, but now for $f_{p_x}$ = 1).
The error bars denote the statistical uncertainty of $\sqrt{N}$ for the 
frequency $N$ in each bin. As we are comparing distributions for two 
different data sets (one at $\phi = 0$, the other at $\phi = \pi/2$), we 
normalize the distributions, as well as the errors, with the total 
number of particles included for each data set. We combine histograms of
500 events in HIJING with the X axis of each event (with  same event 
parameters such as the impact parameter, collision energy
etc.) representing the minor axis of the ellipse representing the overlap
of the two colliding nuclei. We can see from Fig.2a that the two histograms, 
are almost overlapping. This is interesting as we are considering here
non-central collisions (with $b = 3-4$ fm). This shows that the momentum 
distributions and all fluctuations obtained from HIJING (without any 
momentum modification, i.e. $f_{p_x}$ = 1) are isotropic. Any anisotropy
will result from effects of particle interactions, primarily from
collective flow effects, and we will represent it by using values of 
$f_{p_x}$ different from 1.  We mention here that the peak near $0.1$ 
radians in Fig.2 will represent fluctuations of size order 1 fm at the
distance of about 10 fm representing the size of initial nuclei. As 
discussed in ref.\cite{cmbhic}, initial fluctuations are expected because 
of the nucleon size of order 1 fm (as well as quantum fluctuations in
nucleon coordinates.) Such a peak thus may have useful information about 
the scale of fluctuations in the early stages of system evolution.

\begin{figure}
\begin{center}
\leavevmode
\epsfysize=5truecm \vbox{\epsfbox{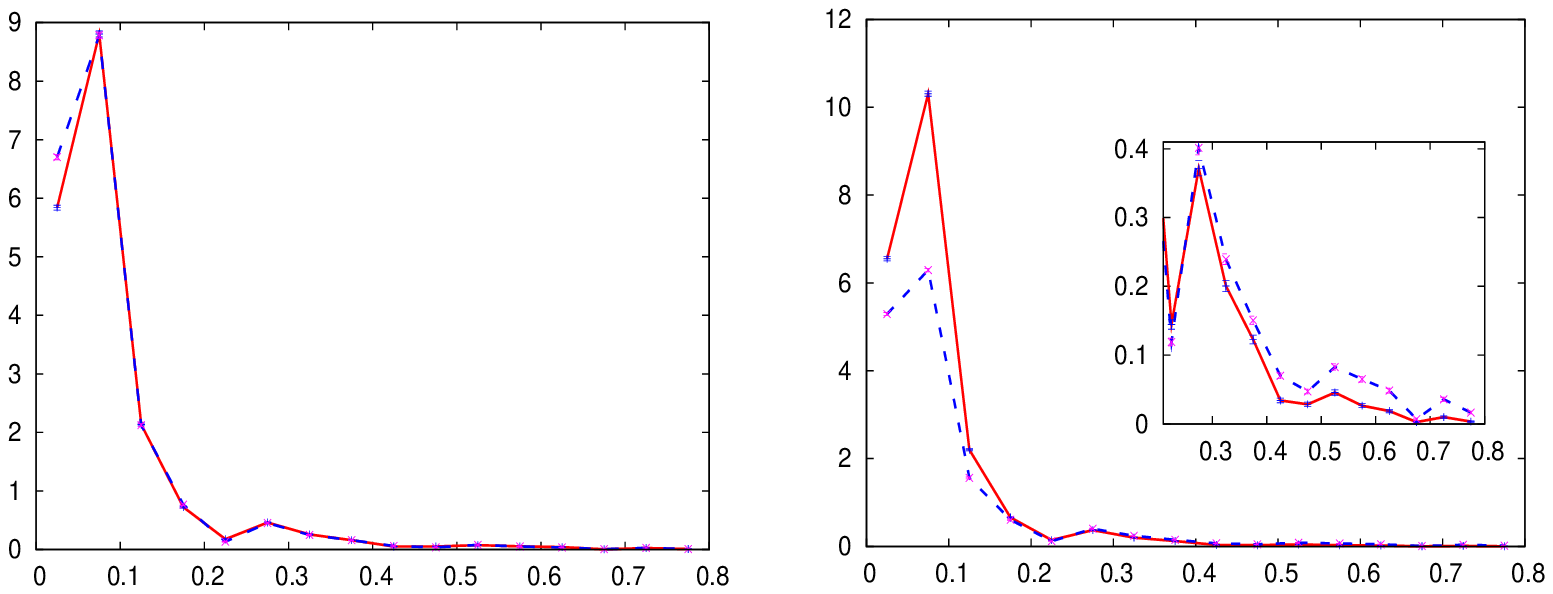}}
\end{center}
\caption{}{(a) Plots of the histograms of the widths (in radians) of filled 
patches for $f_{p_x}$ = 1. Smooth curves join the points which are marked 
with corresponding ($\sqrt{N}$) error bars. Distribution 
for widths at $\phi = 0 (\phi = \pi/2) $ are shown by solid (dashed) 
plots. (b) Plots for the case with momentum modification, with $f_{p_x}$ 
= 1.2. The inset in (b) shows the plots for larger widths.}
\label{Fig2}
\end{figure}

 We now consider an anisotropic momentum distribution by taking
$f_{p_x}$ = 1.2 to modify particle momentum distribution from HIJING.
(Note that this is just to model the effects of flow.) Fig.2b shows the
plots of distributions of widths for this case. We see that the distribution
at $\phi = \pi/2$ is smaller than the distribution at $\phi = 0$
(beyond the error bars) for small widths while the situation is reversed
for larger widths (as shown by the inset in Fig.2b). This is expected as
the stretching of fluctuation patches will be more significant at
$\phi = \pi/2$ due to elliptic flow. Stretching makes fluctuation patches
larger, hence the distribution for large widths is higher for $\phi = \pi/2$
than for $\phi = 0$. This is suitably compensated at smaller angles where
$\phi = \pi/2$ distribution falls below the $\phi = 0$ distribution.
We find that the ratio of the 
heights of the two peaks is directly related to the value of $f_{p_x}$ 
representing the effects of elliptic flow. For $f_{p_x} = $ 1.1, 1.2, 1.3, 
we find this ratio to be 1.3, 1.6, 2.0. Thus, the magnitude of elliptic flow
can be probed directly with the comparison of the distributions of
widths of fluctuation patches at $\phi =  0$ and at $\phi = \pi/2$.
Note that our modeling of flow in terms of a factor like $f_{p_x}$ is
ad hoc. Thus we do not try to derive any functional relationship between
this factor and the ratios of the heights of the two peaks above (or try to 
give error bars for these etc.). Our purpose here is to show specific patterns
of these distributions depending on momentum anisotropy. We also mention 
here that in actual hydrodynamic flow, the effect of flow on the momentum of
particles will depend on the Y coordinate of the particle inside
the plasma (in the event plane), being less important for larger Y values,
as well as on the distance from the center. To more accurately represent the 
$Y$ dependence of flow anisotropy, we have compared the $\phi = 0$ 
distribution using $f_{p_x}$ = 1.3 with the $\phi = \pi/2$ distribution with 
$f_{p_x}$ = 1.1. The difference between the two distributions is found to be 
closer to the difference obtained for the constant value of $f_{p_x}$ = 1.2 
for both the distributions. Since we are only interested in comparative
changes in the two distributions, for simplicity, we will continue to
use a constant value of $f_{p_x}$ for determining the difference between
the $\phi = 0$ and $\phi = \pi/2$ distributions. 

  Of course this requires identification of the event plane so that statistics
of a large number of events can be added up for controlled statistical
errors. We now show that our technique can provide a new way to identify 
the event plane. In Fig.3 , we show a series of plots of widths distributions 
at $\phi = \phi_0$ (solid plot) and at $\phi = \phi_0 + \pi/2$ (dashed curve)
for the angle $\phi_0 = 0, \pi/16, \pi/8, \pi/4$. We note that the ratio of
the two peaks (solid curve and dashed curve) is largest for $\phi_0 = 0$
which represents the event plane. Thus, one needs to simply plot the two 
width distributions at angles $\pi/2$ apart and keep changing the initial
angle until the two distributions have maximum  difference. This will
identify the event plane. (Though we note that maximum overlap of the two
distributions seems to happen at $\phi_0 = \pi/8$ and not at $\pi/4$
as one would have expected. This could be due to the large arc size of 
45$^\circ$ used for calculating the individual distributions.) We have 
also compared the two distributions by varying the separation of $\phi$ at 
which they are evaluated. When the separation is larger than about 
50-60$^\circ$ the difference becomes almost similar up to 90$^\circ$ 
separation. With the physics of ellptic flow in view, we choose 90$^\circ$
separated distributions for comparison. Fig.3 is shown with collection of 
data from 500 events each. One will need to do this on event-by-event basis to 
be able to identify the event plane in each case. Fig.4 shows the same plots 
as in Fig.3, but now for single events. We see that statistical errors are
larger. For larger collision
energies, statistical errors will be under better control. Also, one may 
arrive at the event plane by an iterative process. First one can roughly 
identify event planes for each event (as in Fig.4). Then a large number of 
events can be combined for each choice of $\phi_0$  to maximize the difference
between the two plots. This can be done by systematically changing the
event plane for each event, until the all-event-sum histograms (as in
Fig.3) lead to maximum difference between the two histograms. 

\begin{figure}
\begin{center}
\leavevmode
\epsfysize=8truecm \vbox{\epsfbox{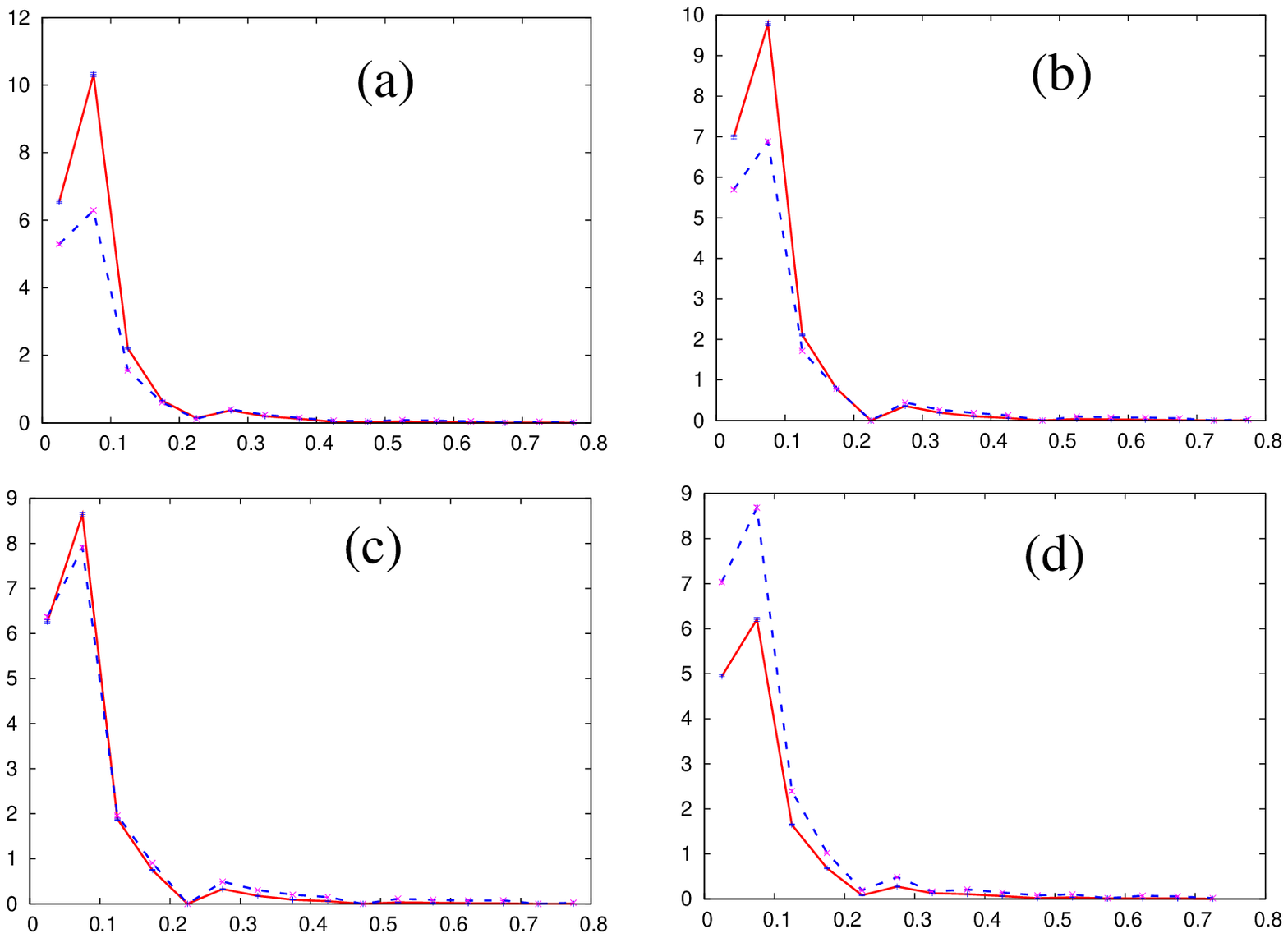}}
\end{center}
\caption{}{Plots of histograms (for 500 events each) at $\phi = \phi_0$ 
(solid curve) and at $\phi_0 + \pi/2$ (dashed curve) for $\phi_0 = 0$ (a), 
$\pi/16$ (b), $\pi/8$ (c), and $ \pi/4$ (d).}
\label{Fig3}
\end{figure}

\begin{figure}
\begin{center}
\leavevmode
\epsfysize=8truecm \vbox{\epsfbox{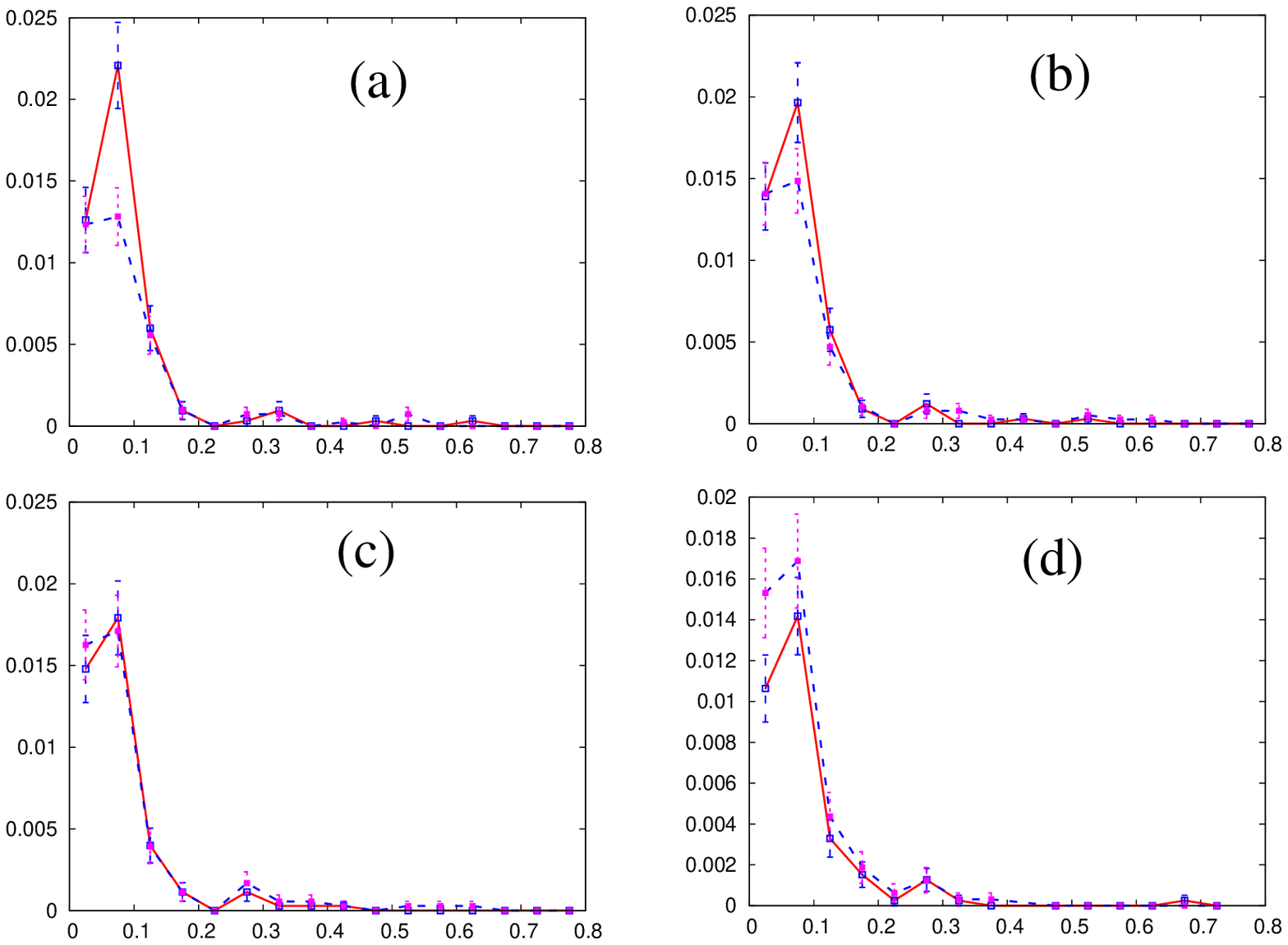}}
\end{center}
\caption{}{Series of plots as in Fig.3, for single event case
each.}
\label{Fig.4}
\end{figure}

In this paper we have shown that a shape/size analysis of excursion sets can
lead to a completely new way of identifying the event plane in RHICE and
to quantify the magnitude of flow anisotropies. This method is novel, and 
completely independent of other conventional methods. This technique can, 
thus, be used to cross-check the results obtained by other conventional 
techniques. Further, the entire distribution of sizes of excursion sets, as
shown in Fig.2, will contain valuable information about nature of fluctuations 
and that of anisotropic flow. Also the width distribution along the rapidity
should provide information about the longitudinal scaling of fluctuations. 
These issues can be probed using hydrodynamic simulations.
Work is underway along these line and we hope to present it in future.
We mention here that  anisotropic expansion of the universe was discussed
in ref.\cite{expn} using Fourier transform technique as well as direct shape
analysis of excursion sets. In this paper we have used the direct shape 
analysis method for investigating flow anisotropy. It will be useful
to also use Fourier transform technique for this purpose. Note that
this technique will detect any effects present in the system which lead to 
anisotropy of momenta. For example the presence of initial strong magnetic 
field (as well as induced electric field) will lead to anisotropies 
\cite{mhd,tuchin} and should leave imprints on these distributions of 
widths. Even the presence of any initial anisotropies of fluctuations, such 
as for deformed nuclei, could be detected in this method. Important issue
then will be to distinguish any initial anisotropy (as from a deformed
nucleus) from anisotropy resulting from anisotropic flow. In this
respect this technique of direct shape analysis seems particularly
effective, in comparison to, e.g. the Fourier transform technique,
as discussed in detail in ref.\cite{expn}. We hope to address these issues
in a future work. We point out 
that other methods such as Minkowski fucntionals \cite{mnkwsk}, analysis 
of genus statistics of excursion sets \cite{praba} etc., have also been 
very effectively used for analyzing morphology of fluctuations in CMBR
as well as in large scale structure in the universe, and it will be
fruitful to apply these techniques also for analyzing the nature of 
fluctuations as well as flow anisotropies in RHICE. 

  We are very grateful to Abhishek Atreya, Partha Bagchi, and Anjishnu 
Sarkar for useful discussions and comments. 


\end{document}